\RequirePackage{ifpdf}
\ifpdf
\documentclass[cits]{PoS-pdf}
\else
\documentclass[cits]{PoS}
\fi

\usepackage{amsmath,amssymb}
\usepackage{amsxtra,mathrsfs}
\usepackage{multirow,multicol}
\usepackage{graphicx}

\ifpdf
\usepackage{epstopdf}
\else
\fi

\title{Numerical evaluation of NLO multiparton processes}

\ShortTitle{Numerical evaluation of NLO multiparton processes}

\author{S. Becker, D. G\"otz, \speaker{C. Reuschle}, C. Schwan and S. Weinzierl\\
         PRISMA Cluster of Excellence, Institut f\"ur Physik,
         Johannes Gutenberg-Universit\"at Mainz\\
         \email{becker@thep.physik.uni-mainz.de\\
                goetz@uni-mainz.de\\
                reuschle@uni-mainz.de\\
                schwan@uni-mainz.de\\
                stefanw@thep.physik.uni-mainz.de}}

\abstract{
We discuss an algorithm for the numerical evaluation of NLO multiparton processes. 
We focus hereby on the virtual part of the NLO calculation, i.e.\ on evaluating the one-loop integration numerically.
We employ and extend the ideas of the subtraction method to the virtual part and we use
subtraction terms for the soft, collinear and ultraviolet regions, which allows us to evaluate the loop integral numerically in four dimensions.
A second ingredient is a method to deform the integration contour of the loop integration into the complex plane. 
The algorithm is derived on the level of the primitive amplitudes, where we utilise recursive relations to generate the corresponding one-loop off-shell currents. 
We discuss the numerical behavior of the approach and the application to the leading colour contribution in $e^+ e^- \rightarrow n \text{ jets}$, with $n$ up to seven.
}

\FullConference{"Loops and Legs in Quantum Field Theory" 11th DESY Workshop on Elementary Particle Physics \\
                 April 15-20, 2012\\
                 Wernigerode, Germany}

\begin{document}

\newcommand{\sla}{\!\!\!/}
\newcommand{\pslash}{\not{\hbox{\kern-2.3pt $p$}}}
\newcommand{\kslash}{\not{\hbox{\kern-2.3pt $k$}}}
\newcommand{\ga}{\gamma}
\newcommand{\de}{\delta}
\newcommand{\ro}{\rho}
\newcommand{\si}{\sigma}
\newcommand{\al}{\alpha}
\newcommand{\be}{\beta}
\newcommand{\la}{\lambda}
\newcommand{\ka}{\kappa}
\newcommand{\eps}{\varepsilon}
\newcommand{\kbar}{\bar{k}}
\newcommand{\scriptA}{\mathcal{A}}
\newcommand{\scriptG}{\mathcal{G}}
\newcommand{\scriptL}{\mathbb{L}}
\newcommand{\scriptI}{\mathbb{I}}
\newcommand{\scriptK}{\mathbb{K}}
\newcommand{\scriptP}{\mathbb{P}}
\newcommand{\order}{\mathcal{O}}
\newcommand{\re}{\mathfrak{Re}}
\newcommand{\im}{\mathfrak{Im}}
\newcommand{\muv}{\mu_{UV}}
\newcommand{\qbar}{\bar{q}}

\section{Introduction}

Calculating efficiently multiparton QCD amplitudes and collider observables at next-to-leading order (NLO) accuracy is a rather involved task.
The most challenging piece is the virtual part.
We discuss here a numerical approach for the calculation of the virtual part, where we
employ the subtraction method and contour deformation \cite{Becker:2012nk,Becker:2012aq,Becker:2011vg,Becker:2010ng,Assadsolimani:2010ka,Assadsolimani:2009cz,Gong:2008ww,Anastasiou:2007qb,Nagy:2006xy,Nagy:2003qn,Soper:2001hu}. 
In this regard our algorithm is different from the commonly used approaches, based on cut techniques and generalised unitarity or on more traditional Feynman graph 
approaches \cite{Berger:2010zx,Ita:2011wn,Ellis:2009zw,Melia:2010bm,Bevilacqua:2010ve,Bevilacqua:2009zn,Bredenstein:2009aj,Frederix:2010ne,vanHameren:2010cp,Badger:2012pf,Badger:2012pg,Cascioli:2011va}, 
and shows promising features for the implementation in a numerical program. 
The algorithm consists of local subtraction terms to subtract divergences arising from the soft, collinear and ultraviolet (UV) regions of the virtual part, 
which render the integrand finite in the respective regions, and of a method to deform the integration contour of the loop integration into the complex plane 
in order to circumvent the remaining on-shell singularities. 
It works on the level of colour-ordered primitive amplitudes, where we utilise recursive algorithms to compute the corresponding one-loop off-shell currents 
for the bare primitive amplitudes, and is therefore fast and easily implemented.
The numerical loop integration is performed together with the integration over the phase-space of the external particles in one Monte Carlo integration. 
The subtraction terms yield simple results upon analytic integration over the loop-momentum, 
and the resulting pole structures cancel exactly against the pole structures from the soft and collinear parts of the real emission contributions as well as 
of the UV counterterm from renormalisation. 
The algorithm goes hand in hand with the usual subtraction method for the real emission contributions, 
where we employ Catani-Seymour dipole subtraction \cite{Catani:1997vz,Dittmaier:1999mb,Phaf:2001gc,Catani:2002hc,Weinzierl:2005dd,Gotz:2012zz}. 
We applied the method to $e^+ e^- \rightarrow n \text{ jets}$, with $n$ up to seven, 
in the large-$N_c$ limit.
Up to four jets we reproduce the known results for the respective jet rates with very good agreement. 
Increasing the number of jets up to 7 shows a good scaling behaviour in CPU time with respect to the number of final state particles.

\section{Subtraction method}

The contributions to an infrared-safe observable at next-to-leading order with $n$ final state particles can be written in a condensed notation as
\begin{align}
\langle O \rangle^{NLO} = \int\limits_{n+1}O_{n+1}d\sigma^{R}+\int\limits_{n}O_{n}d\sigma^{V}+\int\limits_{n}O_{n}d\sigma^{C},
\end{align}
where $d\sigma^{R}$ denotes the real emission contribution, which corresponds to the square of the Born amplitude with $(n+3)$ partons $|\scriptA_{n+3}^{(0)}|^{2}$, 
$d\sigma^{V}$ denotes the virtual contribution, which corresponds to the interference term of the renormalised one-loop amplitude 
with the Born amplitude $2\re(\scriptA_{n+2}^{(0)^{*}}\scriptA_{n+2}^{(1)})$, and $d\sigma^{C}$ subtracts initial state collinear singularities. Each term is separately divergent and only their sum is finite. One adds and subtracts suitably chosen pieces to be able to perform the phase-space integrations and the loop integration by Monte Carlo methods:
\begin{align}
\langle O \rangle^{NLO} &=
\int\limits_{n+1}\left(O_{n+1}d\sigma^{R}-O_{n}d\sigma^{A}\right)\;\;\;+\!\!
\int\limits_{n+\mathrm{loop}}\left(O_{n}d\sigma_{\mathrm{bare}}^{V}-O_{n}d\sigma^{L}\right) \notag\\&+
\int\limits_{n}\Big(O_{n}d\sigma_{CT}^{V}\;\;+\;\;O_{n}\int\limits_{\mathrm{loop}} d\sigma^{L}\;\;+\;\;O_{n}\int\limits_{1} d\sigma^{A}\;\;+\;\;O_{n}d\sigma^{C}\Big),
\end{align}
where we write the renormalised virtual piece as the sum of the bare part and a counterterm part
\begin{align}
\int\limits_{n} O_{n}d\sigma^V = \int\limits_{n} O_{n} \int\limits_{\mathrm{loop}} d\sigma^V_{\mathrm{bare}} + \int\limits_{n} O_{n}d\sigma^V_{CT}.
\end{align}
The first term $\left(O_{n+1}d\sigma^{R}-O_{n}d\sigma^{A}\right)$ is by construction integrable over the $\left(n+1\right)$-particle phase-space and can be evaluated numerically. In the dipole subtraction the subtraction term $O_{n}d\sigma^{A}$ is given by a sum over dipoles. 
The second term $\left(O_{n}d\sigma_{\mathrm{bare}}^{V}-O_{n}d\sigma^{L}\right)$ is by construction integrable over the $n$-particle phase-space and the loop-momentum space in four dimensions. 
We thus extend the subtraction method to the virtual part such that we can evaluate the integral of the one-loop amplitude numerically. 
In any Monte Carlo integration the error scales independently of the dimensionality of the integration region. 
The phase-space and loop integrals can thus be evaluated in a combined Monte Carlo integration 
and no additional costs to evaluate the loop integral separately per phase-space point are introduced. 
After analytical integration of the subtraction terms over the unresolved one-parton phase-space and the loop-momentum space respectively, 
the IR poles of the virtual subtraction term cancel with the IR poles of the real subtraction term and the initial state collinear subtraction term, 
whereas the UV poles of the virtual subtraction term cancel with the UV poles of the counterterm. 
Therefore the third term is also finite and can be evaluated numerically.
In short, the NLO contribution is given as the sum of three finite contributions
\begin{align}
\langle O \rangle^{NLO} &=
\langle O \rangle^{NLO}_{\mathrm{real}}+\langle O \rangle^{NLO}_{\mathrm{virtual}}+\langle O \rangle^{NLO}_{\mathrm{insertion}}.
\end{align}

\section{Colour decomposition and kinematical setup}

Amplitudes in QCD may be decomposed into group-theoretical colour factors multiplied by purely kinematical partial amplitudes:
\begin{align}
\scriptA^{(1)} = \sum_{i} C_iA^{(1)}_i
\end{align}
In the colour-flow basis \cite{Weinzierl:2005dd,'tHooft:1973jz,Maltoni:2002mq} the colour structures are linear combinations of monomials in Kronecker $\delta_{ij}$'s. 
One-loop partial amplitudes are further expressed in terms of linear combinations of primitive one-loop amplitudes, 
where a primitive amplitude is defined as a gauge-invariant set of colour-stripped Feynman diagrams with a fixed cyclic ordering of the external partons 
and a definite routing of the external fermion lines through the diagrams \cite{Bern:1994fz}, which are important properties for our method. 
We will drop any subscripts refering to colour on the primitive amplitudes $A^{(1)}$ from now on.
\begin{figure}
\centering
\includegraphics[scale=1.15]{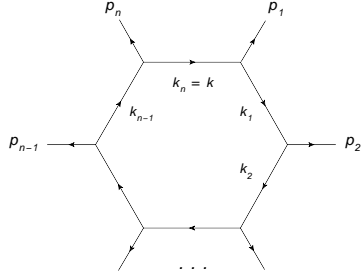}
\caption{\label{fig_kinematics} The labelling of the momenta for a primitive one-loop amplitude.}
\end{figure}
Due to the fixed cycling ordering there are only $n$ different loop-propagators occuring in a primitive amplitude with $n$ external legs. With the notation as in fig.(\ref{fig_kinematics}) we define $k_{j} =  k-q_{j}$, with $ q_{j} =  \sum_{i=1}^{j}p_{i}$, where $k$ denotes the loop-momentum in clock-wise direction, and the $p_{i}$ are the outgoing external momenta. We can write the bare primitive one-loop amplitude, in dimensional regularisation, as
\begin{align}
A_{\mathrm{bare}}^{(1)}=
\int\frac{d^{D}k}{(2\pi)^{D}}G_{\mathrm{bare}}^{(1)}\;\;,\quad\text{ with }\qquad G_{\mathrm{bare}}^{(1)}\ = \ P_{a}(k)\prod\limits_{j=1}^{n}\frac{1}{k_{j}^{2}-m_{j}^{2}+i\delta},
\end{align} 
where $P_{a}(k)$ is a polynomial of degree $a$ in the loop momenta $k$ and the $+i\delta$-prescription tells us in which direction the poles of the propagators should be avoided. Soft singularities arise for $k \sim q_i$ and $p_i^2=m_{i-1}^2$, $m_i=0$, $p_{i+1}^2=m_{i+1}^2$, i.e.\ a massless particle exchanged between two on-shell particles and soft $k_i$. Collinear singularities arise for $k \sim q_i-xp_i$ and $p_i^2=0$, $m_{i-1}=0$, $m_i=0$, where $x \in [0,1]$, i.e.\ a massless external on-shell particle attached to two massless propagators and collinear momenta $k_{i-1}$, $k_i$ and $p_i$. UV singularities arise for $|k|\to\infty$ and $4+a-2n\geq0$.

\section{Local infrared and ultraviolet subtraction terms}

The soft and collinear subtraction terms for massless QCD read
\begin{align}
G_{\mathrm{soft}}^{(1)} &=
4i\sum\limits_{j\in I_g} \frac{p_{j}.p_{j+1}}{k_{j-1}^2k_{j}^2k_{j+1}^2} g^{\mathrm{UV}}_{\mathrm{soft}}(k_{j-1}^2,k_{j}^2,k_{j+1}^2))A_{j}^{(0)}, \\
G_{\mathrm{coll}}^{(1)} &=
-2i\sum\limits_{j\in I_g} \bigg[ \frac{S_{j}g^{\mathrm{UV}}_{\mathrm{coll}}(k_{j-1}^2,k_{j}^2)}{k_{j-1}^2k_{j}^2}
                               + \frac{S_{j+1}g^{\mathrm{UV}}_{\mathrm{coll}}(k_{j}^2,k_{j+1}^2)}{k_{j}^2k_{j+1}^2} \bigg] A_{j}^{(0)},
\end{align}
where the sum over $j \in I_g$ is over all gluon propagators $j$ inside the loop. Furthermore, $S_{j}=1$ if the external line $j$ corresponds to a quark and $S_{j}=1/2$ if it corresponds to a gluon. The functions $g^{\mathrm{UV}}_{\mathrm{soft}}$ and $g^{\mathrm{UV}}_{\mathrm{coll}}$ ensure that the integration over the loop momentum is UV finite. 
They are chosen to be equal to one in the corresponding soft and collinear limits and to suppress the integrands by appropriate powers of $1/|k|$ in the ultraviolet limit $|k|\to\infty$.
The soft and collinear subtraction terms yield simple results upon analytic one-loop integration in dimensional regularisation, 
with the appropriate pole structure to cancel the poles from the real emission and initial state collinear subtraction. 
We derived them for the massless as well as the massive case.

The UV subtraction terms correspond to local counterterms $g^{G}_{\mathrm{UV}}(\kbar,Q,\{p_j\},\{m_j\})$ to propagator and vertex corrections $G$. 
These local counterterms are used in a recursive approach to build a total UV subtraction term to the bare primitive amplitude. 
The local counterterms are obtained by expanding the relevant loop propagators around a new UV propagator $(\kbar^2-\mu_{\mathrm{UV}}^2)^{-1}$, 
where $\kbar = k - Q$ and for a single propagator we have
\begin{align}
\frac{1}{\left(k-p\right)^2} & =
  \frac{1}{\bar{k}^2-\mu_{\mathrm{UV}}^2}
+ \frac{2\bar{k}\cdot\left(p-Q\right)}{\left(\bar{k}^2-\mu_{\mathrm{UV}}^2\right)^2}
- \frac{\left(p-Q\right)^2+\mu_{\mathrm{UV}}^2}{\left(\bar{k}^2-\mu_{\mathrm{UV}}^2\right)^2}
+ \frac{\left[ 2\bar{k}\cdot\left(p-Q\right)\right]^2}{\left(\bar{k}^2-\mu_{\mathrm{UV}}^2\right)^3}
+ {\cal O}\left(\frac{1}{|\bar{k}|^5}\right).
\end{align}
We can always add finite terms to the subtraction terms. For the local counterterms we choose the finite terms such that the finite parts of the integrated local counterterms are independent of $Q$ and proportional to the pole part, with the same constant of proportionality. The integrated local counterterms assume then the general form
\begin{align}
C^{G,(0)}(\{p_j\},\{m_j\}) \Big(\frac{1}{\eps} - \ln\frac{\mu_{\mathrm{UV}}^2}{\mu^2}\Big) + \order(\eps)
\end{align}
where $C^{G,(0)}(\{p_j\},\{m_j\})$ takes on the typical form of the corresponding renormalised colour-ordered Born level propagator or vertex function. This ensures that the sum of all integrated local counterterms is again proportional to a tree-level amplitude.
We derived the local UV counterterms for the massless as well as the massive case.

\section{Contour deformation}

Having a complete list of local UV and IR subtraction terms at hand, we can ensure that the integration over the loop-momentum gives a finite result and can therefore be performed in four dimensions. However, there is still the possibility that some of the loop-propagators go on-shell for real values of the loop-momentum. Therefore we shift the integration contour into the complex space $\mathbb{C}^{4}$, where the integration contour must be chosen such that whenever possible the poles of the propagators are avoided. We set
\begin{align}
k=\tilde{k}+\imath\kappa\left(\tilde{k}\right),
\end{align}   
where $\tilde{k}^{\mu}$ contains only real components. After the deformation our one-loop integral reads
\begin{align}
I=
\int\frac{d^{4}k}{(2\pi)^{4}}
\frac{R(k)}{\prod\limits_{j=1}^{n}\left(k_{j}^{2}-m_{j}^{2}\right)}
=
\int\frac{d^{4}\tilde{k}}{(2\pi)^{4}}\left|\frac{\partial k^{\mu}}{\partial \tilde{k}^{\nu}}\right|
\frac{R(k(\tilde{k}))}{\prod\limits_{j=1}^{n}\left(\tilde{k}_{j}^{2}-m_{j}^{2}-\kappa^{2}+2\imath\tilde{k}_{j}\cdot\kappa\right)}.
\end{align}
where we integrate over the four real components in $\tilde{k}^{\mu}$. To match Feynman's $+\imath\delta$-prescription we have to construct the deformation vector $\kappa$ such that $\tilde{k}_{j}\cdot\kappa \geq 0$ whenever $\tilde{k}_{j}^{2}-m_{j}^{2}=0$,
and the equal sign applies only if the contour is pinched. If the contour is pinched the singularity is either integrable by itself or there is a subtraction term for it. The numerator function $R(k)$ has only poles at
$\bar{k}^2-\mu_{\mathrm{UV}}^{2}=0$.
Choosing $\mu_{\mathrm{UV}}^{2}$ sufficiently large on the negative imaginary axis ensures that the integration contour always stays away from these poles. 
The form of the deformation vector $\kappa$ has a direct impact on the Monte Carlo integration  error.
We employ several techniques to reduce the Monte Carlo integration error.
First, we split our integral into an exterior and an interior part:
\begin{align}
\int \frac{d^4k}{(2\pi)^4} f(k) = \int \frac{d^4k}{(2\pi)^4} f_{\mathrm{UV}}f(k) +  \int\frac{d^4k}{(2\pi)^4} (1-f_{\mathrm{UV}})f(k),
\qquad
f_{\mathrm{UV}}=\prod\limits_{j=1}^{n}\frac{k_{j}^{2}-m_{j}^{2}}{\bar{k}^{2}-\mu_{\mathrm{UV}}^{2}}.
\end{align}
This splitting is holomorphic in $k$, which means we can use different contours for the evaluation of the two parts.
In the exterior part $I_{ext}$ the poles from $k_{j}^{2}-m_{j}^{2}$ are absent and the deformation vector $\kappa$ can simply be chosen as
$\kappa^{\mu} =g_{\mu\nu}\left(\tilde{k}^{\nu}-Q^{\nu}\right)$.
If we choose the arbitrary four-vector $Q$ in $\bar{k}=k-Q$ to be
\begin{align}
Q=\frac{1}{n}\sum\limits_{j=1}^{n}q_{j}
\label{eq:Q}
\end{align} 
we can arrange that the integrand of the interior part drops off with two extra powers in $1/|k|$ for $|k|\to\infty$ and thus receives less contributions from the UV region. The construction of the deformation vector for the internal region is along the lines of \cite{Gong:2008ww} and given in detail in \cite{Becker:2012aq}.

\begin{figure}
\begin{minipage}[t]{0.515\textwidth}
\centering
\includegraphics[viewport=165 600 515 735,scale=0.8]{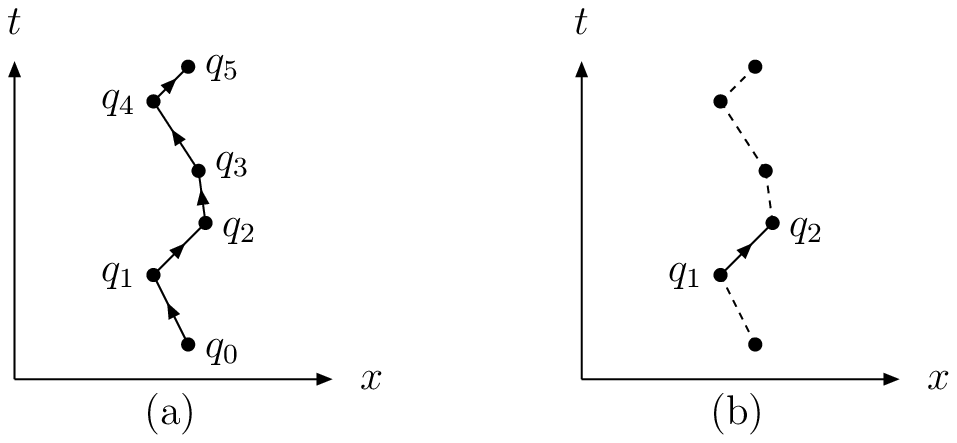}
\caption{Diagram $(a)$ shows the origins of the light cones in electron-positron annihilation. The origins are given by the $(n-1)$ vertices. The vertices are connected by $(n-2)$ line segments. We decompose $I_{int}$ into $(n-2)$ sub-channels, such that each sub-channel corresponds to one line segment. This is shown in diagram $(b)$, where the line segment from $q_1$ to $q_2$ is emphasised.}
\label{fig_zig_zag2}
\end{minipage}
\hspace{2.5mm}
\begin{minipage}[t]{0.485\textwidth}
\centering
\includegraphics[scale=0.4]{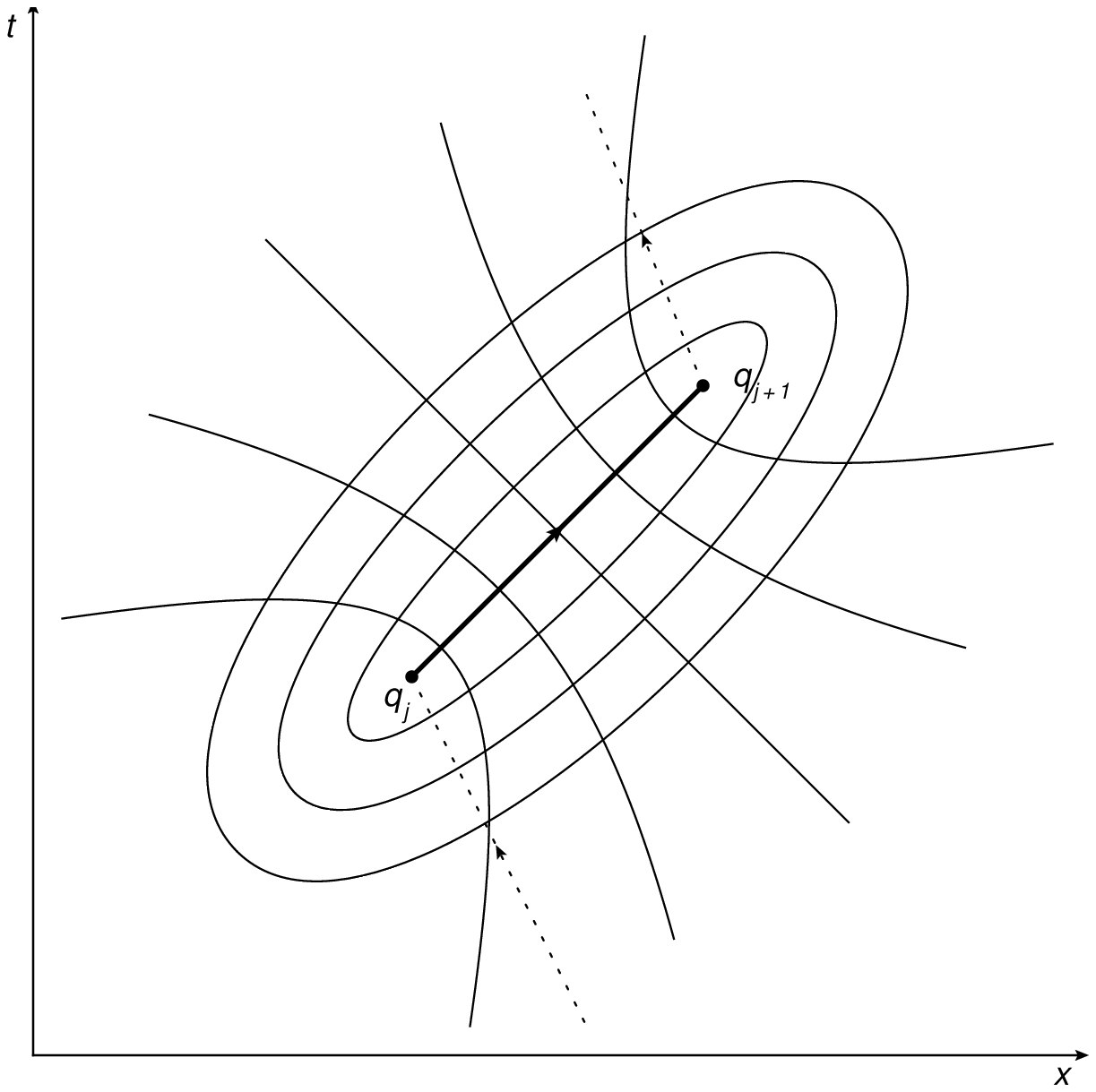}
\caption{The diagram shows the coordinate system for the $j$'th sub-channel. We use  generalisations of elliptical coordinates to four dimensions. At the origin the radial component of the loop-momentum in the elliptical parametrisation equals zero and the loop-momentum coincides with a point on the line segment that connects $q_{j}$ with $q_{j+1}$.}
\label{elliptic1}
\end{minipage}
\end{figure}

The lines connecting the origins of the light cones given by $(k-q_j)=0$ in fig.(\ref{fig_zig_zag2}) are the regions of collinear singularities. If the loop-momentum approaches a collinear singularity we have a subtraction term for it. To improve the numerical behaviour in the vicinity of such a collinear singularity we further 
split the interior part into several sub-channels, where each sub-channel corresponds to one line segment in fig.(\ref{fig_zig_zag2}). We write
\begin{align}
\int\frac{d^{4}k}{(2\pi)^{4}} f(k) = \sum\limits_{j=1}^{n-1}\int\frac{d^{4}k}{(2\pi)^{4}}w_{i}(k)f(k)
\end{align}     
with
$w_{i} \geq 0$ and $\sum\limits_{i=1}^{n-1} w_{i}(k) = 1$.
For the weights $w_{i}$ we use
\begin{align}
w_{i}(k)=\frac{\left(|k_{i}^{2}||k_{i+1}^{2}|\right)^{\alpha}}{\sum\limits_{j=1}^{n-1}\left(|k_{j}^{2}||k_{j+1}^{2}|\right)^{\alpha}}
\end{align}
with $\alpha = -2$. 
This division into sub-channels is not holomorphic and we have to use the same integration contour in all the sub-channels. 
However, we can use different parametrisations, i.e.\ coordinate systems, for the different sub-channels. 
A possible choice to parametrise a certain sub-channel is the generalisation of elliptical coordinates to four dimensions. 
The coordinate system of the sub-channel $j$ is thereby chosen such that the origin of this coordinate system corresponds to the line segment, 
which connects $q_{j}$ with $q_{j+1}$. This is shown in fig.(\ref{elliptic1}). 
The sampling for an individual sub-channel is discussed in detail in \cite{Becker:2012aq}. 
Upon this parametrisation we observe certain periodic oscillations of the integrand in the generalised elliptical coordinates. 
In order to average out these oscillations we use the method of antithetic variates, 
as described in \cite{Becker:2012aq}, which improves the numerical behaviour significantly.

\section{Numerical stability in the UV regions}

It turns out that also the UV regions give rise to large numerical oscillations. 
In their standard form the local UV counterterms are constructed such that after subtraction we are left with a leading $1/|\kbar|^5$-behaviour for $|\kbar|\to\infty$, 
which is formally enough to ensure UV finiteness. 
In order to improve the damping in the UV region, we choose the expansion of the UV denominators such that we are left with a leading $1/|\kbar|^7$-behaviour, 
where our experience shows that we only need to do this for the corrections to propagators and three-valent vertices. 
We do the same for the UV-suppression functions $g_{\mathrm{soft}}^{\mathrm{UV}}$ and $g_{\mathrm{coll}}^{\mathrm{UV}}$ in the soft and collinear subtraction terms. 
Together with the choice for $Q$ in eq.(\ref{eq:Q}) the UV-behaviour of the integrand is significantly enhanced. 
More details and additional comments can be found in \cite{Becker:2012aq}.

\section{Recursion relations}

\begin{figure}
\centering
\begin{align*}
\raisebox{-0.6cm}{\includegraphics[scale=1.25]{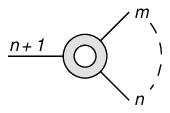}} \quad=\quad
\sum\limits_{i=m}^{n-1}\;\; \raisebox{-0.75cm}{\includegraphics[scale=1.0]{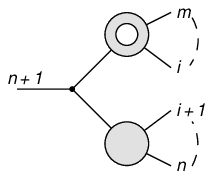}} \quad+\quad
\sum\limits_{i=m}^{n-1}\;\; \raisebox{-0.75cm}{\includegraphics[scale=1.0]{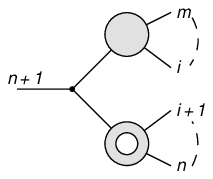}} \quad+\quad
\raisebox{-0.6cm}{\includegraphics[scale=0.6]{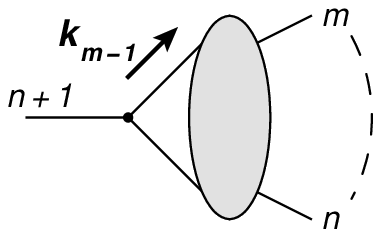}}
\end{align*}
\vspace{-0.25cm}
\begin{align*}
\raisebox{-0.6cm}{\includegraphics[scale=1.25]{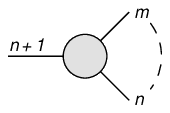}} \;\;=\;\;
\sum\limits_{i=m}^{n-1}\; \raisebox{-0.75cm}{\includegraphics[scale=1.0]{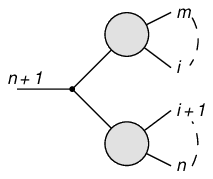}} \qquad\qquad\qquad
\raisebox{-0.75cm}{\includegraphics[scale=1.1]{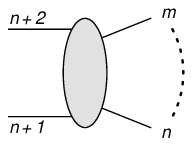}} \;=\;
\sum\limits_{i=m-1}^{n-1}\; \raisebox{-1.1cm}{\includegraphics[scale=1.2]{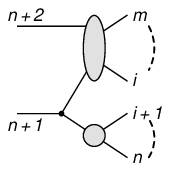}}
\end{align*}
\vspace{-0.5cm}
\begin{align*}
\raisebox{-0.6cm}{\includegraphics[scale=1.25]{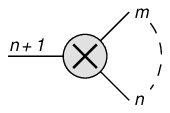}} =
\sum\limits_{i=m}^{n-1} \left[
\raisebox{-0.75cm}{\includegraphics[scale=1.0]{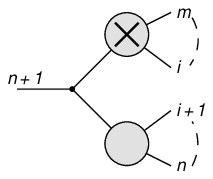}} +
\raisebox{-0.75cm}{\includegraphics[scale=1.0]{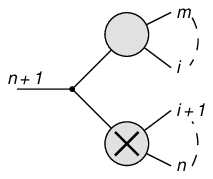}} +
\raisebox{-0.75cm}{\includegraphics[scale=1.0]{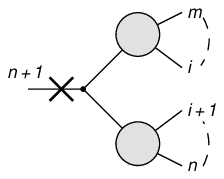}} +
\raisebox{-0.75cm}{\includegraphics[scale=1.0]{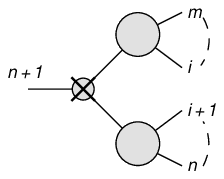}}
\;\right]
\end{align*}
\caption{\label{fig_recursion} Recursive relations for a three-valent toy model.}
\end{figure}

\begin{figure}
\centering
\includegraphics*[viewport=0 10 360 227,scale=0.6]{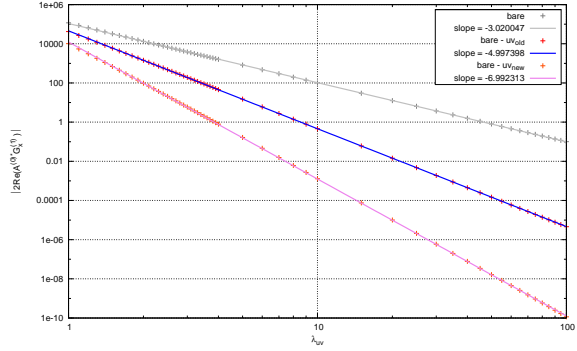}
\caption{\label{uvscaling} The plot shows the helicity summed $|2\re(A^{(0)*} G^{(1)})|$ vs. a UV scaling parameter $\lambda_{\mathrm{UV}}$ 
for $e^+e^- \rightarrow 3 \text{ jets}$, in leading colour approximation, where we scale a fixed value for the loop-momentum according 
to $\kbar = k-Q_{\mathrm{fixed}} = \lambda_{\mathrm{UV}}\kbar_{\mathrm{fixed}}$. 
The unsubtracted total integrand (upper fit in grey) falls off locally like $1/|\kbar|^3$, which leads to UV divergences upon integration. 
The (standard) UV subtracted total integrand (middle fit in blue) falls off like $1/|\kbar|^5$, which is clearly UV finite. 
The (numerically improved) UV subtracted total integrand (lower fit in pink), 
which contains those local UV counterterms that also subtract the $1/|\kbar|^5$- and $1/|\kbar|^6$-behaviour, falls off like $1/|\kbar|^7$.}
\end{figure}

We use Berends-Giele type recursion relations \cite{Berends:1987me} to compute the tree amplitude, the bare one-loop integrand $G^{(1)}_{\mathrm{bare}}$ 
and the total UV subtraction term $G^{(1)}_{\mathrm{UV}}$. 
These recursion relations are shown in fig.(\ref{fig_recursion}) for the case of a three-valent toy model. 
The recursive relations have been implemented for all necessary cases, i.e.\ gluon currents as well as quark and antiquark currents, 
that are needed for $e^+ e^- \rightarrow n \text{ jets}$ in the large-$N_c$ limit. 
In the direct loop contributions, where the off-shell leg couples directly through a vertex to the loop, 
the two edges of the vertex are connected to loop-propagators. 
We can cut open one of these two edges by replacing the tensor structure of the corresponding propagator by a sum over (pseudo)-polarisations.
More details can be found in \cite{Becker:2012aq}. 
We can check the cancellations in the UV region between the recursive constructions of the bare one-loop integrand and the total UV subtraction term. 
This is shown in fig.(\ref{uvscaling}), where we plot $|2\re(A^{(0)*} G^{(1)})|$ vs. a UV scaling parameter $\lambda_{\mathrm{UV}}$ for $e^+e^- \rightarrow 3 \text{ jets}$, in leading colour approximation.

\section{NLO results for $n$ jets in electron-positron annihilation}

\begin{figure}
\begin{minipage}{0.515\textwidth}
\centering
\begin{tabular}{|c|c|c|}\hline
\parbox[0pt][2.5em][c]{0cm}{}$n$ & $N_c\big(\tfrac{N_c}{2}\big)^{n-2}A_{n,lc}$ & $N_c\big(\tfrac{N_c}{2}\big)^{n-1}B_{n,lc}$ \\\hline
\parbox[0pt][2.5em][c]{0cm}{}$5$ & $( 2.4764 \pm 0.0002 ) \cdot 10^{4}$ & $( 1.84 \pm 0.15 ) \cdot 10^{6}$ \\\hline
\parbox[0pt][2.5em][c]{0cm}{}$6$ & $( 2.874 \pm 0.002 ) \cdot 10^{5}$ & $( 3.88 \pm 0.18 ) \cdot 10^{7}$ \\\hline
\parbox[0pt][2.5em][c]{0cm}{}$7$ & $( 2.49 \pm 0.08 ) \cdot 10^{6}$ & $( 5.4 \pm 0.3 ) \cdot 10^{8}$ \\\hline
\end{tabular}
\caption{The results for the LO and NLO leading-colour jet rate coefficients, for a jet parameter of $y_{cut}=0.0006$. Shown are the results for five, six and seven jets.}
\label{fig_table}
\end{minipage}
\hspace{2.5mm}
\begin{minipage}{0.485\textwidth}
\centering
\includegraphics[viewport= 140 460 490 710,scale=0.5]{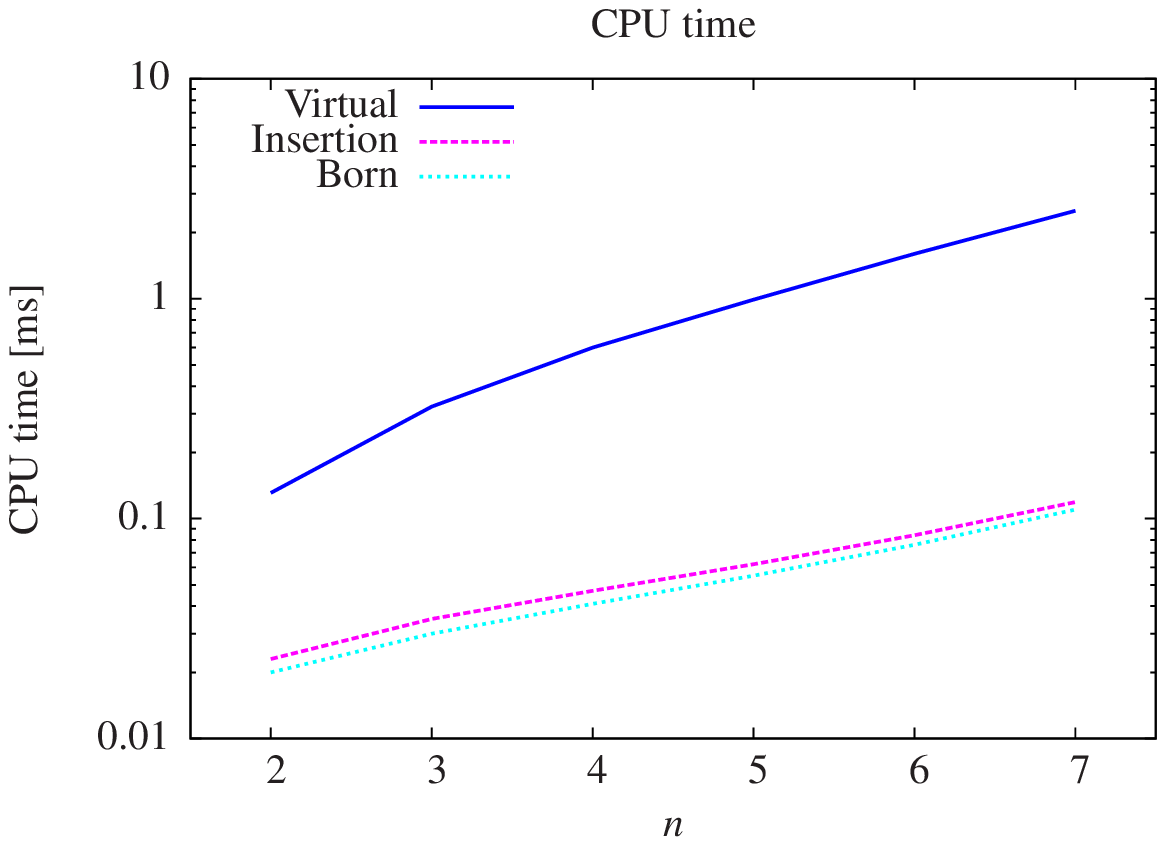}
\caption{CPU time required for one evaluation of the various contributions as a function of the number $n$ of jets. The times are taken on a single core of a standard PC.}
\label{fig_cpu}
\end{minipage}
\end{figure}

We have calculated the Durham jet rates in electron-positron annihilation in the leading colour approximation for up to seven jets \cite{Becker:2011vg}. 
The cross section for $n$ jets normalised to the leading order (LO) cross section for $e^{+}e^{-}\rightarrow$ hadrons reads
\begin{eqnarray}
\frac{\sigma_{n-jet}(\mu)}{\sigma_{0}(\mu)}&=& \left(\frac{\alpha_{s}(\mu)}{2\pi}\right)^{n-2}A_{n}(\mu)+\left(\frac{\alpha_{s}(\mu)}{2\pi}\right)^{n-1}B_{n}(\mu)+\mathcal{O}(\alpha_{s}^{n}).
\end{eqnarray}
One can expand the perturbative coefficient $A_{n}$ and $B_{n}$ in $1/N_{c}$:
\begin{eqnarray}
A_{n}&=& N_{c}\left(\frac{N_{c}}{2}\right)^{n-2}\left[A_{n,lc}+\mathcal{O}\left(\frac{1}{N_{c}}\right)\right],\qquad B_{n}\ = \ N_{c}\left(\frac{N_{c}}{2}\right)^{n-1}\left[B_{n,lc}+\mathcal{O}\left(\frac{1}{N_{c}}\right)\right].
\end{eqnarray}
We calculate the LO coefficient $A_{n,lc}$ and the NLO coefficient $B_{n,lc}$ for $n\leq 7$ at the renormalisation scale $\mu$ equal to the centre of mass energy. We take the centre of mass energy to be equal to the mass of the $Z$-boson. The scale variation can be restored from the renormalisation group equation. The calculation is done with five massless quark flavours.
\begin{figure}
\centering
\includegraphics[viewport= 125 460 490 710,width=0.32\textwidth]{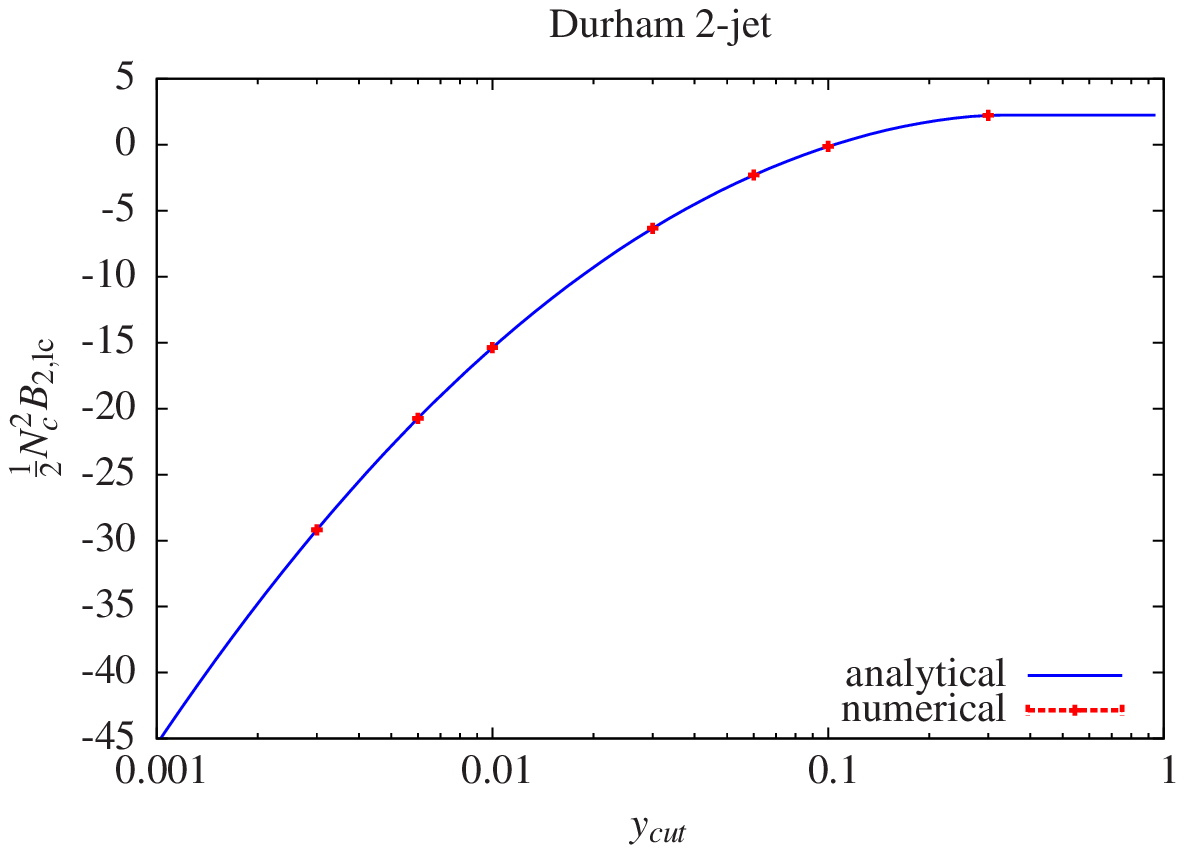}
\includegraphics[viewport= 125 460 490 710,width=0.32\textwidth]{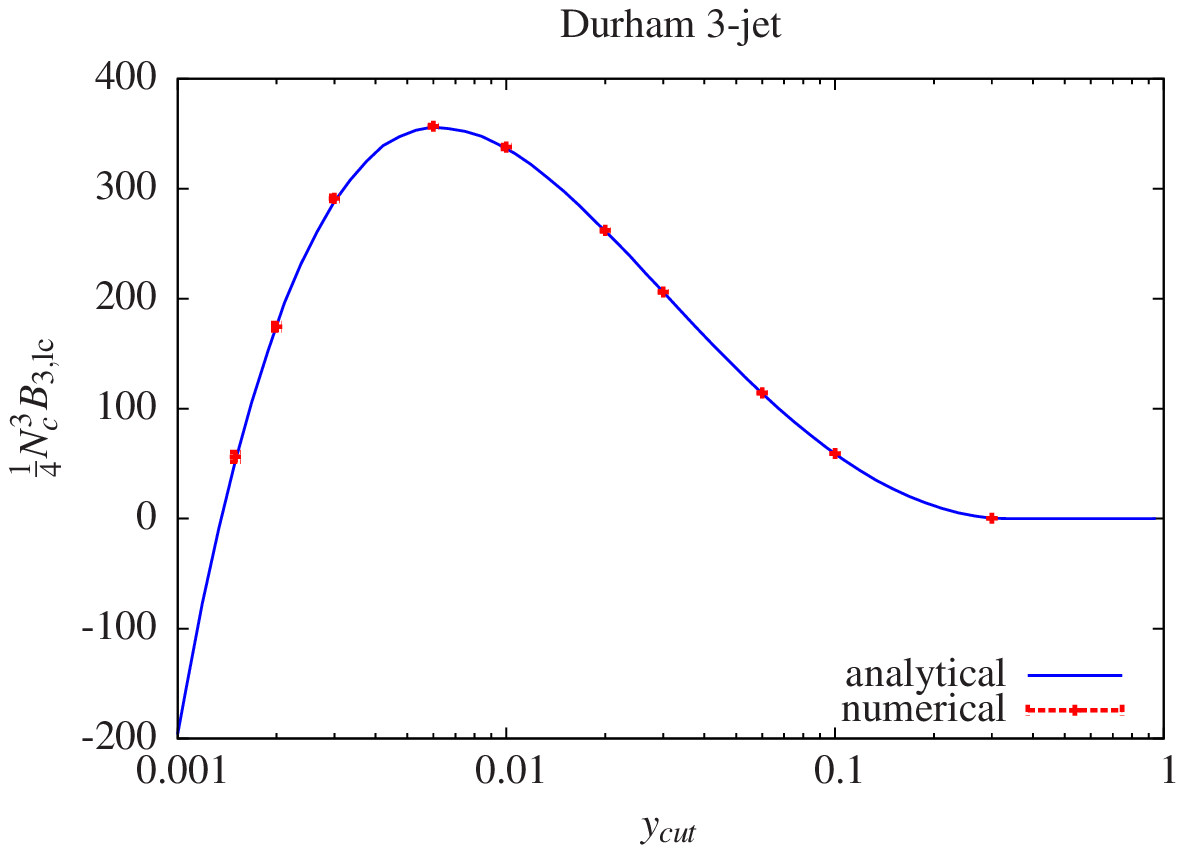}
\includegraphics[viewport= 125 460 490 710,width=0.32\textwidth]{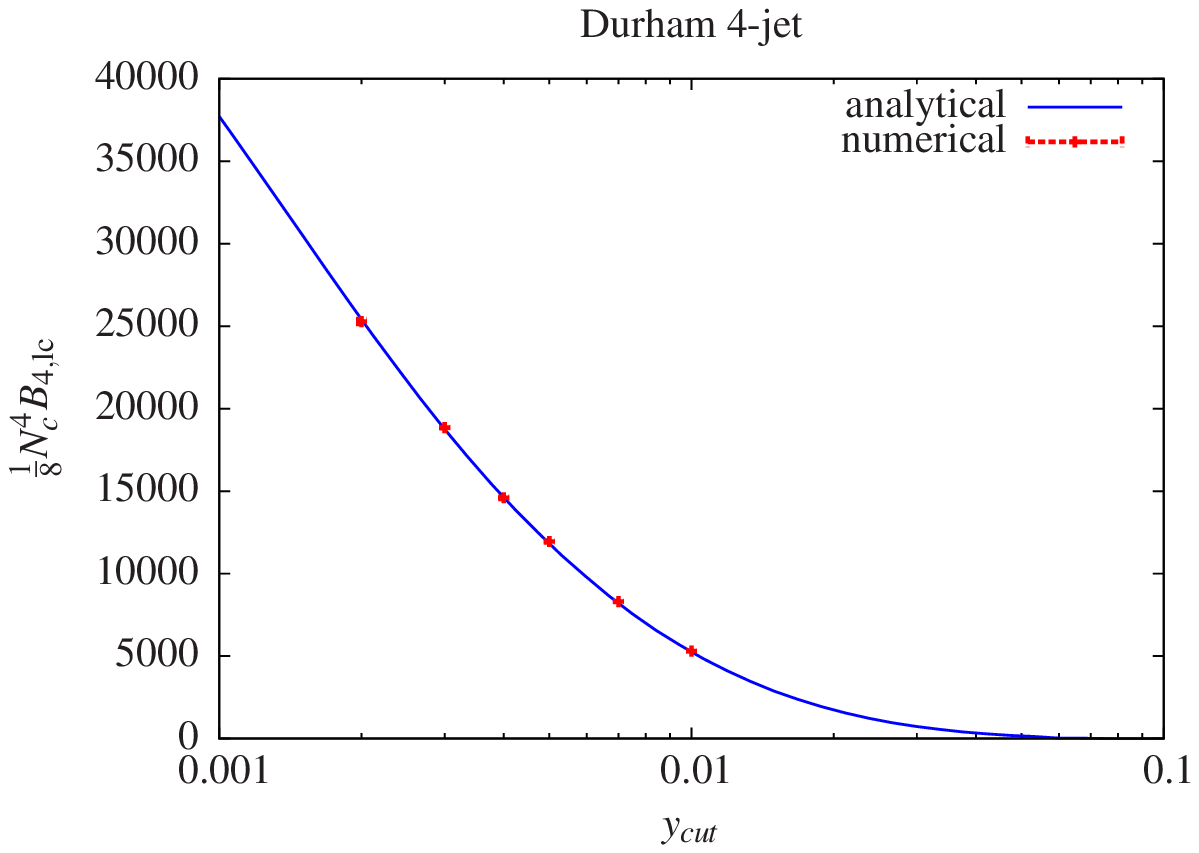}
\caption{\label{fig_jetrates} Comparison of the NLO corrections to the two-, three- and four-jet rate between the numerical calculation and an analytic calculation. The error bars from the MC integration are shown and are almost invisible.}
\end{figure}
Fig.(\ref{fig_jetrates}) shows the comparison of our numerical approach with the well-known results for two, three and four jets \cite{Weinzierl:1999yf,Weinzierl:2010cw}. We observe an excellent agreement. The results for five, six and seven jets for a jet parameter of $y_{cut}=0.0006$ are shown in fig.(\ref{fig_table}).
Fig.(\ref{fig_cpu}) shows the CPU time required for one evaluation of the Born contribution, the insertion term and the virtual term as a function of the number $n$ of jets. One notes that the insertion term is almost as fast as the Born contribution. For all contributions the CPU time per evaluation increases only very moderately as a function of $n$. Within our method all three contributions scale asymptotically as $n^4$, as expected \cite{Kleiss:1988ne}. The virtual part has the same scaling behaviour as the Born contribution. This moderate growth imposes almost no restrictions on the number of final state partons to which our method can be applied. The practical limitations arise from the fact that the number of evaluations required to reach a certain accuracy increases with $n$. This behaviour is already present at the Born level and not inherent to our method. The calculation of the seven-jet rate takes about five days on a cluster with 200 cores.

\bibliography{/home/stefanw/notes/biblio}
\bibliographystyle{/home/stefanw/latex-style/h-physrev5}

\end{document}